\documentclass[proceedings]{JHEP3} 

\usepackage{epsfig}                  
\usepackage{graphicx}

\newcommand{\TM} {\texttt{TM}}
\newcommand{\TMs} {\texttt{TM}s}
\newcommand{\CL} {\% C.L.}
\newcommand{\la}{|\mathbf{\Lambda}|}

\PrHEP{}
\conference{International Workshop on Astroparticle and High Energy Physics}

\title{Constraining neutrino magnetic moment with solar and reactor neutrino data}

\author{\speaker{M.~A.~T\'ortola} \thanks{In collaboration with  
  W.~Grimus, M.~Maltoni, T.~Schwetz and J.~W.~F.~Valle}
  \\
  Instituto de F\'{\i}sica Corpuscular --
  C.S.I.C./Universitat de Val{\`e}ncia \\
  Edificio Institutos de Paterna, Apt 22085,
  E--46071 Valencia, Spain.\\
  E-mail: \email{mariam@ific.uv.es}}

\abstract{  
  We use solar neutrino data to derive stringent bounds on Majorana neutrino
  transition moments (\TMs). Such moments, if present, would contribute to
  the neutrino--electron scattering cross section and hence alter the signal
  observed in Super-Kamiokande. Using the latest solar neutrino data, combined 
  with the results of the reactor experiment KamLAND, we perform a simultaneous 
  fit of the oscillation parameters and \TMs. Furthermore, we show how the inclusion 
  of data from the reactor experiments Rovno, TEXONO and MUNU in our analysis improves 
  significantly the current constraints on \TMs. Finally, we perform a simulation of 
  the future Borexino experiment and show that it will improve the bounds from today's 
  data by order of magnitude.
  }

\begin{document} 

\section{Introduction}

Present solar~\cite{homestake,sage,GaGNO,superk,SNO} and atmospheric neutrino 
data \cite{atmos} provide the first robust evidence for neutrino flavour conversion 
and, consequently, the first solid indication for new physics. Neutrino oscillations 
constitute the most popular explanation for the data (for recent analysis see
Refs.~\cite{Maltoni:3nu,Maltoni:2002ni,solfits}) and are a natural outcome of gauge 
theories of neutrino mass~\cite{Schechter:1980gr,rev}. Besides neutrino oscillations, 
non-zero neutrino masses manifest themselves also through non-standard neutrino 
electromagnetic properties. In the minimal extension of the Standard Model (SM) 
neutrinos get Dirac masses together with magnetic moments (MMs)~\cite{Fujikawa:yx}, 
although these MMs are too small to see any effect at current experiments.
However, if we consider some extensions beyond the SM, neutrinos may have MMs of order 
$10^{-10}$--$10^{-11} \mu_B$, relevant for the present sensitivities.
In particular, if we assume that neutrinos are Majorana particles, then the 
structure of their electromagnetic properties is characterized by a $3 \times
3$ complex antisymmetric matrix, the so-called Majorana transition moment (\TM) 
matrix, that contains MMs as well as electric dipole moments of the 
neutrinos~\cite{Schechter:1981hw}.

\section{Theoretical framework}
\label{sec:theor-fram}

In this work (based on Ref.~\cite{grimus}), we consider that only three light 
neutrinos exist, which is well-motivated by recent global fits of all available 
neutrino oscillation data~\cite{Maltoni:2002xd}. We restrict our analysis to the 
LMA-MSW solution of the solar neutrino problem, as indicated by recent global fits 
of solar data~\cite{Maltoni:3nu,Maltoni:2002ni,solfits} and confirmed by the reactor
experiment KamLAND~\cite{kamland}. On the other hand, we assume that neutrinos are 
endowed with \TMs\ and have Majorana nature, as expected from 
theory~\cite{rev,Schechter:1981hw}.

In experiments where the neutrino detection reaction is elastic neutrino--electron 
scattering, like in Super-Kamiokande (SK), Borexino and some reactor
experiments~\cite{rovno,texono,munu}, the electromagnetic cross section 
is~\cite{Bardin:wr}
\begin{equation}\label{cross}
\frac{d\sigma_\mathrm{em}}{dT} = \frac{\alpha^2 \pi}{m_e^2 \mu_B^2}
\left(\frac{1}{T} - \frac{1}{E_\nu}\right) \mu^2_\mathrm{eff} \,,
\end{equation}
where the effective MM square is given by~\cite{GS}
\begin{equation}\label{mmeff}
\mu^2_\mathrm{eff} =  
a_-^\dagger \lambda^\dagger \lambda a_- +
a_+^\dagger \lambda \lambda^\dagger a_+
\,.
\end{equation}
The electromagnetic cross section adds to the weak cross section and allows
to extract information on the \TM\ matrix $\lambda$. In this cross section,
$T$ denotes the kinetic energy of the recoil electron and $E_\nu$ is the energy
of the incoming neutrino. According to Eq.~(\ref{cross}), one notices that the 
electromagnetic contribution to the total cross section is more important at low 
energies and therefore experiments observing neutrinos with lower energies will be 
more sensitive to \TMs. 
The 3-vectors $a_-$ and $a_+$ denote the neutrino amplitudes for negative and 
positive helicities at the detector. The square of the effective MM given in 
Eq.~(\ref{mmeff}) is independent of the basis chosen for the neutrino state~\cite{GS}. 
In what follows we consider both the flavour basis and the mass eigenstate basis. 
We will use the convention that $a_\mp$ and $\lambda = (\lambda_{\alpha\beta})$
denote the quantities in the flavour basis, whereas in the mass basis we will use 
$\tilde a_\mp$ and $\tilde\lambda =(\lambda_{jk})$.  The two bases are connected via 
the neutrino mixing matrix $U$:
\begin{equation}\label{trafo}
\tilde a_- = U^\dagger a_- \,, \quad
\tilde a_+ = U^T       a_+ \,, \quad
\tilde \lambda = U^T \lambda U \,.
\end{equation}
Taking into account the antisymmetry of the transition moment matrix $\lambda$ for 
Majorana neutrinos, it is useful to define vectors $\mathbf\Lambda = (\Lambda_\alpha)$ 
and $\tilde{\mathbf\Lambda} = (\Lambda_j)$ in the flavour and mass basis,
respectively, by
\begin{equation}\label{defL}
\lambda_{\alpha\beta} = \varepsilon_{\alpha\beta\gamma} \Lambda_\gamma
\quad \mathrm{and} \quad 
\lambda_{jk} = \varepsilon_{jkl} \Lambda_l \,.
\end{equation}
Thus, in the flavour basis we have $\lambda_{e \mu} = \Lambda_\tau$,
$\lambda_{\mu\tau} = \Lambda_e$ and $\lambda_{\tau e} = \Lambda_\mu$.
Note also that 
\begin{equation}
\label{basisindep}
\la^2 = \frac{1}{2}\, \mathrm{Tr} \left( \lambda^\dagger \lambda \right)
\quad \Rightarrow \quad
\la = |\tilde{\mathbf\Lambda}| \,.
\end{equation}
This means that, if we are able to find a bound on $\la$, we have not only 
constrained the \TMs\ in the flavour basis but also in the mass basis. 

Now we will present the form that the effective MM square $\mu_\mathrm{eff}^2$ in 
Eq.~(\ref{mmeff}) takes in the cases of solar and reactor neutrino experiments. 
The detailed derivation of the following expressions can be found in
Ref.~\cite{grimus}. In the context of the LMA-MSW solution of the solar neutrino 
problem, we obtain the effective MM square
\begin{equation}\label{finalLMA}
\mu^2_\mathrm{LMA} = \la^2 - |\Lambda_2|^2 +
P^{2\nu}_{e1} \left( |\Lambda_2|^2 - |\Lambda_1|^2 \right) \,.
\end{equation}
Here $\Lambda_i$ are the components of the \TM\ matrix in the mass basis, and 
$P^{2\nu}_{e1}$ corresponds to the probability that an electron neutrino produced 
in the core of the sun arrives at the detector as the mass eigenstate $\nu_1$ in 
a 2--neutrino scheme. 

In the case of reactor neutrinos,  the $\mu_\mathrm{eff}^2$ relevant in reactor 
experiments is given as
\begin{equation}\label{reactorMM}
\mu_\mathrm{R}^2 = |\Lambda_\mu|^2 + |\Lambda_\tau|^2 \,.
\end{equation}
From this relation it is clear that reactor data on its own cannot constrain all 
\TMs\ contained in $\lambda$, since $\Lambda_e$ does not enter in 
Eq.~(\ref{reactorMM}). In order to combine reactor and solar data it is useful to 
rewrite $\mu_\mathrm{R}^2$ in terms of the mass basis quantities
\begin{equation} \label{finalreactor}
\mu_\mathrm{R}^2 = 
\la^2 - c^2 |\Lambda_1|^2 - s^2 |\Lambda_2|^2 
-2 s c |\Lambda_1| |\Lambda_2| \cos\delta \,,
\end{equation}
where $c=\cos\theta$ and $s=\sin\theta$, $\theta$ being the solar mixing angle. 
Note that the relative phase $\delta=\mathrm{arg}(\Lambda_1^* \Lambda_2)$ between 
$\Lambda_1$ and $\Lambda_2$ appears in addition to $\la$, $|\Lambda_1|$ and
$|\Lambda_2|$.

\section{Statistical Method}
\label{sec:stat}

In the following we will use data from solar and reactor neutrino experiments to 
constrain neutrino \TMs. The $\chi^2$-function obtained from the data depends on 
the solar oscillation parameters $\tan^2\theta$ and $\Delta m^2$ as well as on 
the elements of the \TM\ matrix $\lambda$.  Regarding the dependence on the oscillation
parameters we will take two different attitudes. One is to assume that $\tan^2\theta$ 
and $\Delta m^2$ will be determined with good accuracy at the KamLAND experiment, 
and hence, we will consider the $\chi^2$ at fixed points in the 
$\tan^2\theta - \Delta m^2$ plane (method I). In the second approach we will derive 
a bound on the \TMs\ by minimizing $\chi^2$ with respect to $\tan^2\theta$ and 
$\Delta m^2$ (method II). This second procedure takes into account the present
uncertainty of our knowledge of the oscillation parameters.

Let us describe in detail how we calculate a bound on the \TMs. Our aim is to 
constrain $\la$, therefore it is convenient to consider the $\chi^2$ as a function 
of the independent parameters $\la,|\Lambda_1|,|\Lambda_2|$ and $\delta$. 
The $\chi^2$-functions which we are using for the individual data sets
(solar rates, SK recoil electron spectrum, reactor data) will be described in detail 
in the following sections.  When performing the fit to the data, we find that 
in general the minimum of the $\chi^2$ occurs close or outside the physical boundary 
of the parameters $\la,|\Lambda_1|,|\Lambda_2|$. To take this into account we apply
Bayesian techniques to calculate an upper bound on $\la$. We minimize the $\chi^2$ 
with respect to $|\Lambda_1|$, $|\Lambda_2|$ and $\delta$ for each value of $\la$, 
taking into account the allowed region $0 \le |\Lambda_1|^2 + |\Lambda_2|^2 \le \la^2$:
\begin{equation}\label{minL}
\chi^2(\la)= \mathrm{Min}  \left[
\chi^2(\la,|\Lambda_1|,|\Lambda_2|,\delta) \right] \,.
\end{equation}
In method I we do this for fixed values of the oscillation parameters, scanning 
over the LMA-MSW region, whereas in method II we minimize also with respect to 
$\tan^2\theta$ and $\Delta m^2$ in Eq.~(\ref{minL}). Then the $\chi^2$ is transformed 
into a likelihood function via
\begin{equation}
\mathcal{L} \propto \exp\left( -\frac{1}{2} \chi^2 \right) \,.
\end{equation}
Now we use Bayes' theorem and a flat prior distribution $p(\la)$ in the physically 
allowed region, $p(\la)=\Theta(\la)$, to obtain a probability distribution for $\la$:
\begin{equation}
f(\la) \, d\la= \frac{\mathcal{L}(\la) \, \Theta(\la) \, d\la }
{\int_0^\infty \mathcal{L}(\la') \, d\la' } \,.
\end{equation}
An upper bound $b_\alpha$ on $\la$ at a C.L.\ $\alpha$ is given by the
equation
\begin{equation}
\int_0^{b_\alpha} f(\la)d\la = \alpha \,.
\end{equation}

Analyzing the minimization of Eq.~(\ref{minL}) in more detail (see Ref.~\cite{grimus}), 
we realize that the bound on $\la$ is strongest if $P_{e1}=0.5$ because in this 
situation $\mu_\mathrm{LMA}^2$ in Eq.~(\ref{finalLMA}) is maximal. In 
Fig.~\ref{fig:Pe1} we show contours of constant $P_{e1}$ in the 
$\tan^2\theta - \Delta m^2/E_\nu$ plane.  For definiteness, the probabilities in 
the figure are obtained by performing the averaging over the production distribution 
inside the sun for the $^7$Be flux most relevant for Borexino. However, the 
probabilities for the $^8$B flux relevant for SK are very similar.

\FIGURE[t] {
\includegraphics[width=0.6\linewidth,height=0.45\textheight]{probab.eps}
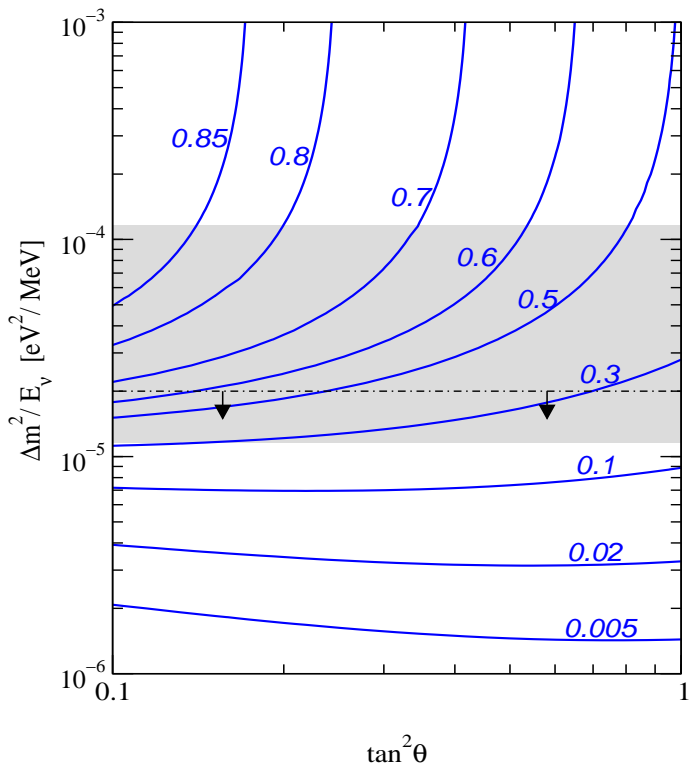
  \caption{Contours of constant $P_{e1}$. The shaded region is relevant for Borexino, 
    whereas the region below the dash-dotted line is relevant for SK.}
  \label{fig:Pe1}
  }
  
The shaded region in Fig.~\ref{fig:Pe1} is the region relevant for
Borexino, assuming a mass-squared difference in the range $10^{-5}$
eV$^2 < \Delta m^2 < 10^{-4}$ eV$^2$, whereas for SK the region below
the dash-dotted line is most important, due to the higher energy of
the $^8$B neutrinos. We can read off from the figure that in most part
of the SK region $P_{e1}$ is very small, which means that the sensitivity of SK 
for $\la$ is limited. In contrast, we expect a much better sensitivity of Borexino 
because, in a large part of the relevant parameter space in this case, $P_{e1}$ 
is close to the optimal value of $0.5$.

\section{Analysis of solar and reactor neutrino data}
\label{sec:analys-solar-react}

In this section we describe briefly our analysis of solar neutrino data and the 
data which we are using from reactor neutrino experiments. If we neglect 
the electromagnetic contribution to the small elastic scattering
signal in the Sudbury Neutrino Observatory (SNO), 
the only solar neutrino experiment whose signal will be affected by neutrino \TMs\
is SK. However, the uncertainty in the $^8$B flux leads to correlations between
SK and all other experiments. Therefore, also other experiments give some information 
on \TMs\ and it is important to include them in the analysis~\cite{joshipura}.

We divide the solar neutrino data into the total rates observed in all
experiments and the SK recoil electron spectrum (with free overall
normalization). Subsequently we consider the reactor data in combination with the
global sample of solar data.

\subsection{Solar rates}

We include in our fit the event rates measured in the chlorine
experiment Homestake \cite{homestake}, the gallium experiments
Sage \cite{sage}, Gallex and GNO \cite{GaGNO}, the total event rate of
SK \cite{superk} based on the 1496 days data sample and the result of 
SNO
\cite{SNO} for the charged current and neutral
current solar neutrino rates. For the analysis we use the $\chi^2$-function
\begin{equation}\label{ratechi2}
\chi^2_\mathrm{rates} = \sum_{j,k} 
(R_j - D_j) (V^\mathrm{rate}_{jk})^{-1} (R_k - D_k) \,.
\end{equation}
Here the indices $j,k$ run over the 6 solar neutrino rates, $D_j$ are the 
experimental rates and $R_j$ are the theoretical predictions, which
are calculated as described in Ref.~\cite{Maltoni:2002ni}, except for 
the SK experiment, whose rate includes an extra contribution
from the electromagnetic scattering, as shown in Ref.~\cite{grimus}.
The covariance matrix $V^\mathrm{rate}_{jk}$ takes into account the experimental errors
and theoretical uncertainties from detection cross sections and SSM predictions of 
the neutrino fluxes (for details see Refs.~\cite{Maltoni:2002ni,fogli}).

\subsection{The Super-Kamiokande recoil electron spectrum}

In this section we consider the shape of the SK recoil
electron spectrum. The electromagnetic contribution to the
neutrino--electron scattering cross section leads to a different 
spectrum of the scattered electrons than expected from the SM weak
interaction. Therefore, the SK spectral data provide a useful tool
to constrain neutrino \TMs~\cite{beacom}.

We perform a fit to the latest 1496 days SK data presented in Ref.~\cite{superk} 
as 44 data points $D_i$, using the following $\chi^2$-function
\begin{equation}\label{chi2spect}
\chi^2_\mathrm{spect} =
\sum_{i,j=1}^{44} (\alpha R_i - D_i) (V^\mathrm{spect}_{ij})^{-1} 
(\alpha R_j - D_j) \,.
\end{equation}
The theoretical prediction $R_i$ is calculated as explained in Ref.~\cite{grimus}.
The covariance matrix $V^\mathrm{spect}_{ij}$ contains statistical and 
systematic experimental errors, extracted from Ref.~\cite{superk}. The $\chi^2$
is minimized with respect to the normalization factor $\alpha$ in order to isolate 
only the shape of the spectrum.

\subsection{Reactor data}

Data from neutrino electron scattering at nuclear reactor experiments can be 
used to constrain the combination of \TMs\ given in Eq.~(\ref{finalreactor}).
In fact, these experiments are the most sensitive probe for laboratory
searches of neutrino magnetic moment because of the high flux of
low-energy antineutrinos emitted by a nuclear reactor and the good experimental control 
through the on-off reactor comparison.
Here we use data from the Rovno nuclear power plant~\cite{rovno}, the TEXONO
experiment at the Kuo-Sheng Power Plant in Taiwan~\cite{texono} and
from the MUNU experiment at the Bugey reactor~\cite{munu}.
To include this information in our analysis we make the following ansatz for 
the $\chi^2$-function
\begin{equation}\label{chi2react}
\chi^2_\mathrm{reactor} (\mu_\mathrm{R}) = 
\sum_i
\left( \frac{N^i_\mathrm{weak} + N^i_\mathrm{em}(\mu_\mathrm{R}) - N^i_\mathrm{obs} } 
{\sigma^i} \right)^2 \,.
\end{equation}
The sum is over the three experiments.
$N^i_\mathrm{obs}$ is the observed number of events with the one
standard deviation error $\sigma^i$, $N^i_\mathrm{weak}$ is the number
of events expected in the case of no neutrino \TMs\ (only the standard
weak interaction) and $N^i_\mathrm{em}(\mu_\mathrm{R})$ is the number of events
due to the electromagnetic scattering of neutrinos with an effective
MM $\mu_\mathrm{R}$. 

\section{Bounds on $\la$ from solar and reactor data}
\label{sec:bounds-la-from}

\FIGURE[t] {
  \includegraphics[width=0.6\linewidth]{cnt_sol.eps}
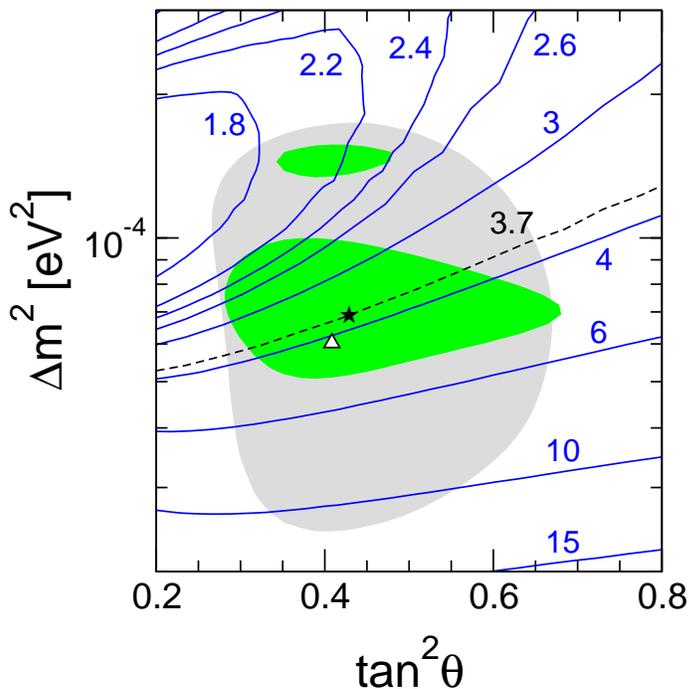
  \caption{Contours of the 90\CL\ bound on $\la$ in units of 
    $10^{-10}\mu_B$ from solar data. The gray (light) shaded region is the 3$\sigma$
    LMA-MSW region obtained in the global analysis of solar neutrino
    data (best fit point marked with a triangle), whereas the green
    (dark) one corresponds to the 3$\sigma$ region obtained after the
    inclusion of the KamLAND results in the analysis (best fit point
    marked with a star), both from Ref.~\cite{Maltoni:3nu}. The dashed
    line shows the bound obtained at the global best fit point.}
  \label{fig:sol}
  }

First we discuss the bounds on $\la$ from solar data alone. Fixing
the oscillation parameters at the current best fit point
$\tan^2\theta=0.43$, $\Delta m^2=6.9\times 10^{-5}$ eV$^2$ from the
global analysis of solar neutrino data including the KamLAND results~\cite{Maltoni:3nu},
we obtain with the Bayesian methods described in
Section~\ref{sec:stat} the 90\CL\ bound
\begin{equation}
\la < 3.7 \times 10^{-10}\mu_B \quad\mbox{(best fit point, solar data)}.
\end{equation}
However, such a bound substantially depends on the values of the neutrino
oscillation parameters. In Fig.~\ref{fig:sol} we show contours
of the 90\CL\ bound on $\la$ in the $\tan^2\theta - \Delta m^2$ plane.
We find that the bound gets weaker for small values of $\Delta m^2$,
whereas in the upper left part of the LMA-MSW region a bound of the order
$2\times 10^{-10}\mu_B$ is obtained.

By combining solar and reactor data we obtain considerably stronger
bounds. At the best fit point we get at 90\CL
\begin{equation}
\la < 1.7 \times 10^{-10}\mu_B \quad
\mbox{(best fit point, solar + reactor data)}.
\end{equation}

\FIGURE[t] {
  \includegraphics[width=0.6\linewidth]{cnt_glob.eps}
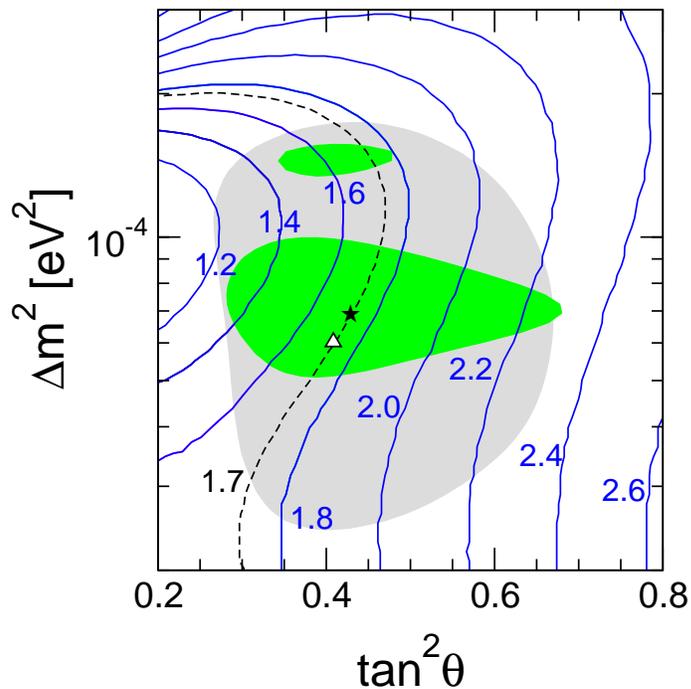
  \caption{Same as Fig.~\ref{fig:sol}, but here the contours refer to the 90\CL\ 
    bound on $\la$ in units of $10^{-10}\mu_B$ from combined solar and
    reactor data.}
  \label{fig:glob}
  }

In Fig.~\ref{fig:glob} we show the contours of the bound in
the $\tan^2\theta - \Delta m^2$ plane for the combination of solar and
reactor data. By comparing Fig.~\ref{fig:sol} with
Fig.~\ref{fig:glob} we find that reactor data drastically
improve the bound for low $\Delta m^2$ values. 

In the upper parts of the LMA-MSW region, the solar data alone give already
a strong bound on $\la$. This is due to the fact that there the probability 
$P_{e1}$ relevant in SK is close to the optimal value of 0.5 (see Fig.~\ref{fig:Pe1}). 
In this parameter region the combination with reactor data does not improve
the bound significantly. In contrast, for low $\Delta m^2$ values the
probability $P_{e1}$ is very small at the neutrino energies relevant
for SK, as seen in Fig.~\ref{fig:Pe1}. Therefore, solar data give
a very weak constraint and only the combination with reactor data improves the bound.

Up to now we have calculated bounds on neutrino \TMs\ for fixed values
of the oscillation parameters $\tan^2\theta$ and $\Delta m^2$ (method
I described in Section~\ref{sec:stat}). These results will be especially
useful after KamLAND will have determined the oscillation parameters
with good accuracy. In the following we change our strategy and
minimize the $\chi^2$ for each value of $\la$ with respect to
$\tan^2\theta$ and $\Delta m^2$ (method II). This will lead to a bound
on $\la$ taking into account the current knowledge concerning the
oscillation parameters. To this end we make use of the global solar neutrino data 
 and the latest KamLAND results~\cite{kamland}, as described in Ref.~\cite{Maltoni-kl} 
in order to obtain the correct $\chi^2$ behavior as a function of
$\tan^2\theta$ and $\Delta m^2$. Performing this analysis we obtain
the following bounds at 90\CL:
\begin{equation}\label{unconstrbounds}
\la < \left\{ \begin{array}{l@{\quad}l}
4.0 \times 10^{-10}\mu_B  & \mbox{(unconstrained, solar + KamLAND data)} \\
1.8 \times 10^{-10}\mu_B  &
\mbox{(unconstrained, solar + KamLAND + reactor data).}
\end{array}\right.
\end{equation}

Together with Figs.~\ref{fig:sol} and \ref{fig:glob},
the bounds given in Eq.~(\ref{unconstrbounds}) show that current solar neutrino 
data can be used to constrain all elements of the Majorana neutrino \TM\
matrix. A combination with data from reactor experiments significantly
improves the bound on $\la$.

\section{Comparison with other works}
\label{sec:compar}

In this section we compare the quality of our analysis with respect to
two previous works doing similar calculations.
In particular, we have chosen the first analysis trying to constrain neutrino 
magnetic moments with solar data, done by Beacom and Vogel~\cite{beacom}, and
a more recent article presented by Joshipura and Mohanty~\cite{joshipura}. 

In Ref.~\cite{beacom}, the SK recoil electron spectrum (825 days) is used 
to constrain the only component of the Majorana \TM\ matrix present in a 2--neutrino 
scheme, $\Lambda_3$. They obtained $|\Lambda_3|$ $\le$ 1.5$\times$10$^{-10}\mu_B$ at 
90\CL\ Using the same assumptions, and just including the SK spectrum shape
(1496 days) in the analysis, we get at 90\CL\ $|\Lambda_3|$ $\le$
1.4$\times$10$^{-10}\mu_B$, which shows the equivalence of the two methods.
Taking advantage of the global analysis presented in this work, using all the available 
information, a better bound is obtained:
$|\Lambda_3|$ $\le$ 7.8$\times$10$^{-11}\mu_B$ at 90\CL\

In the second work~\cite{joshipura}, the authors assume the general case of 
three neutrinos and calculate limits on the three elements different
from zero of the Majorana \TM\ matrix. To do this, they consider just one component 
non-zero at a time, and obtain a constraint for each \TM\ separately in the mass
basis. As data sample they use only the solar neutrino rates. To compare with their 
results, we consider just one \TM\ at a time, but we include  all the data sample in our 
analysis. Results are summarized in Tab.~\ref{tab:compar}, where we can see the
improvement of the present analysis with respect the two works considered.

\TABLE[t]{
\begin{tabular}{|c|c|c|c|}
    \hline
    \multicolumn{4}{|c|}{Bounds at 90\CL}\\
    \hline
    & Beacom \& Vogel~\cite{beacom} & Joshipura \& Mohanty~\cite{joshipura} & 
    present analysis \\
    \hline
    $|\Lambda_1|$ $\le$ & -- & ~4.3$\times$10$^{-10}\mu_B$ &
    1.2$\times$10$^{-10}\mu_B$ \\
    $|\Lambda_2|$ $\le$ & -- & 20.5$\times$10$^{-10}\mu_B$ &
    1.0$\times$10$^{-10}\mu_B$ \\
    $|\Lambda_3|$ $\le$ & ~1.5$\times$10$^{-10}\mu_B$ &~3.8$\times$10$^{-10}\mu_B$ &
    7.8$\times$10$^{-11}\mu_B$ \\
    \hline
\end{tabular}
\caption{Comparison of the results from our work and two previous analysis.}
\label{tab:compar}
}

Finally, we want to remark that our results are the more general
presented up to now, because they are calculated in a
3--neutrino framework, in a basis--independent way and our limits apply to all
components of the \TM\ matrix simultaneously. In addition, we
have used all the available experimental data, and our
bounds are stronger than the obtained in previous calculations.

\section{Sensitivity on $\la$ of the Borexino experiment}
\label{sec:sensitivity-borexino}

Here we investigate the sensitivity of the Borexino
experiment~\cite{borexino} to neutrino \TMs. This
experiment is mainly sensitive to the solar $^7$Be neutrino flux,
which will be measured by elastic neutrino--electron scattering.
Therefore, Borexino is similar to SK, the main difference is the
mono-energetic line of the $^7$Be neutrinos, with an energy of 0.862
MeV, which is roughly one order of magnitude smaller than the energies
of the $^8$B neutrino flux relevant in SK.

To estimate the sensitivity of Borexino we consider the following $\chi^2$-function
\begin{equation}
\chi^2_\mathrm{borexino} =
\sum_{i,j=1}^{N_\mathrm{bins}} 
(N_i^\mathrm{th} - D_i) (V^\mathrm{borex}_{ij})^{-1} (N_j^\mathrm{th} - D_j) \,.
\end{equation}
Here $N_i^\mathrm{th}$ is the theoretical prediction for the number of
events in the electron recoil energy bin $i$, given by the sum of events from
weak scattering, electromagnetic scattering and background
\begin{equation}\label{Nth}
N_i^\mathrm{th} = N_i^\mathrm{weak} +
N_i^\mathrm{em}(\mu_\mathrm{LMA}) + 
N_i^\mathrm{bg} 
\,. 
\end{equation}
$D_i$ is the (hypothetical) observed number of events that, in order to estimate 
the sensitivity of Borexino for neutrino \TMs, we assume generated by neutrinos 
without \TMs
\begin{equation}
D_i = N_i^\mathrm{weak} + N_i^\mathrm{bg} \,.
\end{equation}
Hence, we obtain
\begin{equation}\label{chi2bor}
\chi^2_\mathrm{borexino} =
\sum_{i,j=1}^{N_\mathrm{bins}} 
N_i^\mathrm{em}\,  (V^\mathrm{borex}_{ij})^{-1} \, N_j^\mathrm{em} \,.
\end{equation}
The minimum of this $\chi^2$ occurs for $\mu_\mathrm{LMA}=0$ and is
always zero. With this $\chi^2$ we calculate a
bound on $\la$ at a given C.L. That bound corresponds to the maximum
allowed value of $\la$ which cannot be distinguished from $\la = 0$,
and is therefore a measure for the obtainable sensitivity to $\la$ at
Borexino. The definition of the covariance matrix in Eq.~(\ref{chi2bor}) and
further details about the simulation of Borexino can be found in 
Ref.~\cite{grimus}. 

\FIGURE[t] {
  \includegraphics[width=0.6\linewidth]{cnt_borex.eps}
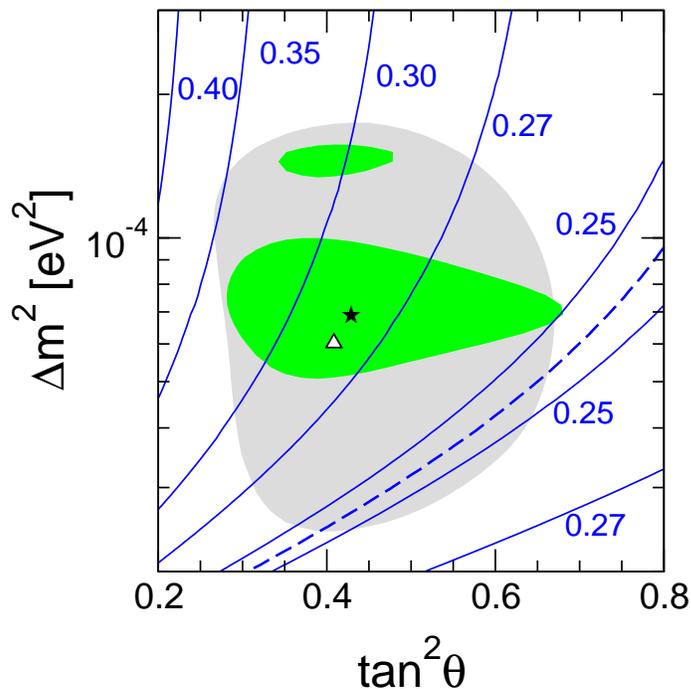
  \caption{Contours of the 90\CL\ bound on $\la$ after 3 years of
    Borexino data-taking in units of $10^{-10}\mu_B$.     
    The gray (light) shaded region is the allowed LMA-MSW region at 
    3$\sigma$ obtained in the global solar analysis (best fit point marked 
    with a triangle), and the green (dark) one corresponds to the 3$\sigma$ 
    region obtained including KamLAND in our analysis (best fit point marked 
    with a star), from Ref.~\cite{Maltoni:3nu}.
    The dashed line corresponds to $P_{e1}=0.5$ for $^7$Be neutrinos,
    and shows the strongest attainable limit.}  
  \label{fig:borex}
  }

\vspace{2mm}

Using the statistical method described previously we obtain for the current best fit point
$\tan^2\theta = 0.43$, $\Delta m^2 = 6.9\times10^{-5}$ eV$^2$ the upper
bound (sensitivity)
\begin{equation}\label{borexinobound}
\la \le 0.28 \times 10^{-10} \mu_B \quad\mbox{at}\quad 90\%\:\mbox{C.L.}
\end{equation}
after three years of Borexino data taking. This bound is about one
order of magnitude stronger than the bound from existing data, which
is partly due to the lower energies observed in Borexino, in
comparison with SK. We have checked that a combined analysis of Borexino with 
existing data (solar and reactor data) does not improve the bound of
Eq.~(\ref{borexinobound}). In Fig.~\ref{fig:borex} we show
contours of the 90\CL\ bound in the $\tan^2\theta-\Delta m^2$ plane.
The strongest attainable limit is roughly $0.24\times 10^{-10} \mu_B$
and, in agreement with the discussion we gave in Sec.~\ref{sec:stat}, 
it is obtained when $P_{e1}=0.5$, as shown by the dashed line in Fig.~\ref{fig:borex}.

\vspace{15mm}

\section{Conclusions}
\label{sec:conclusions}

In this paper we have presented stringent bounds on electromagnetic
Majorana transition moments (\TMs). Such \TMs\ can
contribute to the elastic neutrino--electron scattering signal in
solar neutrino experiments like Super-Kamiokande or the upcoming
Borexino experiment, as well as in reactor experiments that
detect neutrinos through the neutrino--electron elastic scattering
process. Using most recent global solar neutrino data we have
performed a fit in terms of the oscillation parameters and the
elements of the complex \TM\ matrix $\lambda$ of three active Majorana
neutrinos. Taking into account the antisymmetry of the \TM\ matrix, we have
shown that solar neutrino data allow to constrain all Majorana \TMs\
simultaneously in a basis independent way, through the intrinsic
neutrino property $\la$.

A fit to the global solar neutrino data leads to the bound $\la <
4.0\times 10^{-10}\mu_B$ at 90\CL\ We have also considered the role of
reactor neutrino data on neutrino \TMs, shown to be complementary to
solar neutrino data.  A combined fit of reactor and solar data leads
to significantly improved bounds: at 90\CL\ we get $\la < 1.8\times
10^{-10}\mu_B$. 
In the near future a precise determination of the oscillation
parameters might be possible at the KamLAND experiment, which motivated us to scan the
$\tan^2\theta - \Delta m^2$ plane and to calculate the corresponding
bound on $\la$ in each point.  These results are shown in
Figs.~\ref{fig:sol} and \ref{fig:glob} for solar data
and solar + reactor data, respectively.
Comparing our results with previous calculations done by other groups we
have shown that we have obtained the most general and stringent bounds
presented up to this moment.

We have also investigated the potential of the upcoming
neutrino--electron scattering solar experiment Borexino to
constrain neutrino \TMs.  Performing a detailed simulation of the
experiment we find that it will improve the bound on $\la$ by about
one order of magnitude with respect to present bounds.

\acknowledgments
This work has been supported by the Spanish grant BFM2002-00345 and the MECD fellowship 
AP2000-1953.

\end{document}